\documentclass[letterpaper, 10 pt, conference]{ieeeconf}  
\usepackage[]{graphicx}  

\usepackage{xcolor} 
\usepackage{epsfig} 
\usepackage{mathptmx} 
\usepackage{times} 
\usepackage{amsmath} 
\usepackage{amssymb}  
\usepackage{url} 
\usepackage{float}

\title{\LARGE \bf The Solar System as a lab for the Law of Universal Gravitation}

\author{Mauricio Mendivelso-Villaquirán\\ Gimnasio La Montaña\\ Bogotá, Colombia
}

\begin{document}

\maketitle
\thispagestyle{empty}
\pagestyle{empty}

\begin{abstract}
The Law of Universal Gravitation is part of middle and high school's general physics and astronomy curricula. This topic is included in the most popular physics textbooks available as a fact whose origin remains in the detailed work of Sir Isaac Newton 300 years ago. Consequently, its mathematical form is presented as an equation without any deductive process. Nevertheless, deduction of the mathematical form of this law is an opportunity to discuss how a deductive process can be performed using the data available on the Internet from reliable sources. 

\end{abstract}

\section{Introduction}

\noindent The Law of Universal Gravitation (LUG) offers Newton’s elegant solution to the longstanding challenge of describing planetary motion around the Sun. By the 16\textsuperscript{th} century, European observatories had amassed extensive numerical data from celestial observations. Scholars like Johannes Kepler analyzed this data and formulated empirical laws governing planetary motion. While Kepler's laws hinted at a deeper underlying principle, earlier attempts by Galileo and others to generalize them often led to models that were either overly complex or redundant. It was Isaac Newton who succeeded in synthesizing these insights into a unified, coherent theory. His model, in contrast to those built on convoluted trajectories, provided a clear explanation, quickly gaining widespread acceptance.

\bigskip

\noindent At its foundation, LUG describes planetary motion as the result of a gravitational attraction between two masses, with a force proportional to the product of the masses and inversely proportional to the square of the distance between them

\bigskip

\begin{equation}
F = G \frac{m_1 m_2}{r^2} \label{LUG}
\end{equation}

\bigskip

\noindent where \( G \) is the gravitational constant, \( m_1 \) and \( m_2 \) are the masses involved, and \( r \) is the distance separating them. This inverse-square relationship has been extensively tested and confirmed. Although widely featured in general physics textbooks across high school and university levels, the derivation of this law is often absent. As with many physics topics, instructional focus tends to emphasize the final formula and its consequences rather than the path leading to its discovery. Consequently, LUG is frequently presented as an isolated result, historically rooted in Galileo’s time but rarely contextualized in terms of how it was initially deduced.

\bigskip

\noindent Many educational efforts have attempted to make the implications of LUG more approachable. While some of these are reproducible in the classroom, they often leave students wondering how the law was originally derived. When learners ask, ``How can I, as a student, arrive at this result?'' they are commonly referred to post-Galilean or Newtonian interpretations---often found in dated, complex texts---or are simply encouraged to focus on the law’s implications, with little attention to its derivation.

\bigskip

\noindent Surprisingly, there exists a more accessible way to address this question---through the use of a 20\textsuperscript{th}-century tool that transformed our understanding of galactic dynamics: the rotation curve. This straightforward method can be applied to systems governed by gravity---such as the Solar System, satellites in orbit, and even exoplanet systems---to empirically derive LUG. Using rotation curves in tandem with software tools, students can approximate LUG and the Cavendish constant \( G \) with remarkable precision. This approach not only simplifies the derivation process but also reflects the analytical practices of modern science, making the result more tangible and engaging.

\bigskip

\noindent From a pedagogical perspective, this methodology illuminates the origins of LUG and demystifies its development. It also promotes collaborative inquiry and mutual learning as students work together toward a shared scientific goal. Furthermore, the integration of computational tools embodies the principles of STEM education, allowing learners to connect historical breakthroughs with contemporary technology in a meaningful way. In doing so, this activity reinforces gravitational principles while showcasing the relevance and power of STEM-based learning in reconstructing foundational scientific achievements.

\bigskip

\section{LUG in Textbooks and Other Sources}

\noindent Physics textbooks serve as the primary resource for investigating on the presentation of the Law of Universal Gravitation (LUG). Historically and presently, they remain key channels for the dissemination of scientific knowledge. To begin this investigation, a curated list of textbooks was assembled. The motivation behind this compilation stems from the wide variety of available physics textbooks and their differing levels of accessibility. Although using an existing list such as the one by Leite \cite{ref1} was considered advantageous, the age and limited distribution of those titles required an alternative approach.

\bigskip

\noindent In physics education, curricular frameworks frequently rely on textbooks as foundational resources. For example, the Advanced Placement (AP) program, developed by the College Board in the United States, heavily integrates textbooks into its curriculum. Due to the structure of the U.S. educational system, access to textbooks is often driven by demand and market availability, with widely adopted titles enjoying broader circulation. Platforms like Reddit, eBay, and Amazon proved useful for identifying the most accessible and commonly used books. Using this strategy, the textbooks shown in Table~\ref{tab:textbook_lug} emerged as central to this review.

\begin{table}[H]
\centering
\begin{tabular}{|p{2.5cm}|p{3.4cm}|p{0.8cm}|}
\hline
\textbf{Authors} & \textbf{Title} & \textbf{Edition} \\
\hline
E. Etkina, M. Gentile, and A. Van Heuvelen & \textit{College Physics} & 3rd \\ \hline
R. Freedman, T. Ruskell, P. Kesten, and G. Stewart & \textit{College Physics for the AP® Physics 1 \& 2 Courses} & 2nd \\ \hline
D. Giancoli & \textit{Physics} & 7th \\ \hline
D. Halliday, R. Resnick, and J. Walker & \textit{Fundamentals of Physics} & 10th \\ \hline
C. Hamper & \textit{Pearson Baccalaureate: Higher Level Physics} & 3rd \\ \hline
K. Johnson & \textit{Physics for You} & 3rd \\ \hline
T. Kirk & \textit{IB Physics Study Guide: Oxford IB Diploma Programme} & 3rd \\ \hline
R. Knight, B. Jones, and S. Field & \textit{College Physics: A Strategic Approach} & 3rd \\ \hline
R. A. Serway \& J. W. Jewett & \textit{Physics for Scientists and Engineers with Modern Physics} & 10th \\ \hline
K. A. Tsokos & \textit{Physics for the IB Diploma} & 2nd \\ \hline
P. P. Urone, R. Hinrichs & \textit{OpenStax College Physics 2e} (online textbook) & Jul 13, 2022 \\ \hline
H. C. Verma & \textit{Concepts of Physics (Part 1)} & 1st \\ \hline
H. D. Young \& R. A. Freedman & \textit{University Physics with Modern Physics} & 15th \\
\hline
\end{tabular}
\caption{University, college, and AP/IB/GCSE general physics textbooks under review}
\label{tab:textbook_lug}
\end{table}

\noindent Additional frameworks such as the International Baccalaureate Organization’s (IBO) International Baccalaureate program and the British General Certificate of Secondary Education (GCSE) have also been included. These resources offer complementary perspectives on how LUG is presented across educational systems. Although not exhaustive, this selection reflects textbooks commonly used in physics instruction in recent years.

\bigskip

\noindent After reviewing these textbooks, a consistent pattern emerges: LUG is typically introduced in its mathematical form, accompanied by explanations of the equation's components and its implications. Most authors highlight the law’s universality and general applicability without addressing its origin or derivation. At the university level, some texts introduce LUG by invoking Kepler’s Third Law and the nearly circular nature of planetary orbits. This leads to a derivation of the proportionality between the square of the orbital period and the cube of the mean orbital radius, with \( G \) introduced as a proportionality constant. While valid, this method often feels artificial or disconnected, leaving the core question posed in the introduction—how students can deduce LUG—unanswered.

\bigskip

\noindent Turning to other resources, some articles \cite{ref3,ref4,ref5,ref6} examine the implications of LUG's derivation, often citing the measurement of the Cavendish Constant \( G \), or detailing methods used to validate its value. From a historical standpoint, various research groups have studied how the deduction of \( G \) is represented in educational literature \cite{ref7}.

\bigskip

\noindent Videos are also useful learning tools, though their content is often influenced by popularity metrics, which may affect depth and rigor. A non-exhaustive online search revealed several videos in which LUG is either derived from Kepler’s Third Law or presented directly using its standard mathematical form. Some videos also attempt to calculate \( G \) using handmade Cavendish balance models, but they generally begin with the assumption of LUG’s validity, rather than deriving it. Additionally, various scientific enthusiasts have reproduced Cavendish’s results with commendable precision. However, two challenges arise when considering these methods for classroom use. First, the experimental setups are often delicate and require careful handling. Second, the gravitational constant \( G \) is typically assumed rather than derived, raising further questions—many of them philosophical—that the experimenters do not explore.

\bigskip

\noindent Thus, the central question remains open: How can students independently deduce the Law of Universal Gravitation in a meaningful way, rather than relying solely on its pre-existing form as found in textbooks or replicated experiments?

\bigskip

\section{Deduction of LUG}

\noindent Observations of the Solar System have provided us with a wealth of both calculated and observed data on planets and satellites. The Space Race further enriched this body of knowledge by supplying new data from space missions exploring distant bodies within our solar system and beyond. Over the years, this growing dataset has been compiled, refined, and made accessible through public and reliable repositories, such as the NASA Solar System Data service \cite{ref2}.

\bigskip

\noindent There is something particularly compelling in this data. Turning briefly to the work of galactic astronomers, one finds a powerful tool used to study galactic dynamics: the rotation curve. Astronomers like Vera Rubin employed rotation curves during the 1960s \cite{ref9} to analyze the rotational behavior of spiral galaxies. Applying the same logic to our Solar System\footnote{Planetary and satellite systems, such as those referenced here, are essentially the same type of system as a galaxy. Both can be modeled using the classical gravitational potential, which proves sufficiently relevant in both contexts. In light of this, the rotation curve model and the consequences derived from its application successfully account for the rotation of satellites and planets as expected. This underscores its relevance in the description of such systems.}, one can use the orbital velocities and mean distances\footnote{As the orbit around the central body deviates from a perfect circle, the semi-major and semi-minor axes are averaged to approximate the orbital radius or distance.} to the Sun to generate a rotation curve of the planetary system. By using the NASA data, the rotation curve for the Solar System can be plotted. 

\begin{figure}[H]
    \centering
    \includegraphics[width=0.35\textwidth]{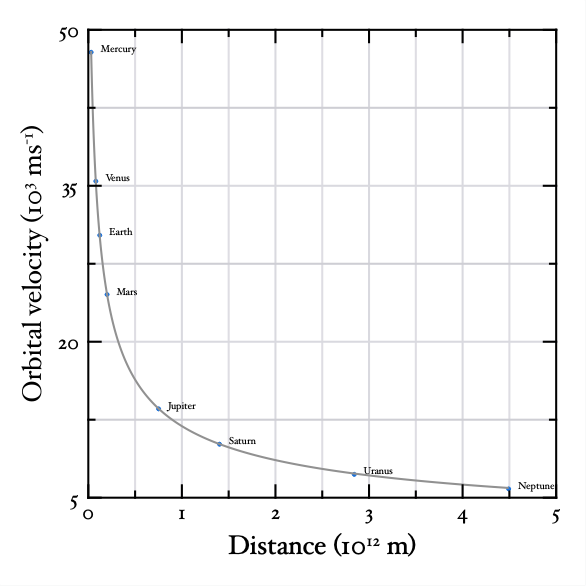}
    \caption{The Solar System rotation curve.}
    \label{fig:rotationcurves}
\end{figure}

\noindent In addition to the Sun-centered system, gravitational systems such as Jupiter, Saturn, Neptune, and Uranus can be analyzed, as their satellites are included in the NASA datasets. \footnote{There are certain considerations related to the intrinsic error involved in the curve fitting process that merit attention. Nevertheless, given that such analysis lies beyond the scope of this proposal, it is important to note that the fitting procedure follows the least squares method, which ensures the minimization of the associated error to the greatest extent possible.}

\begin{figure}[H]
    \centering
    \includegraphics[width=0.35\textwidth]{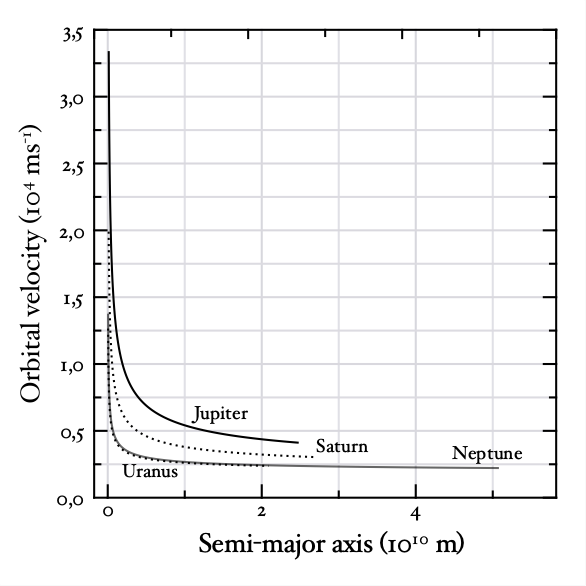}
    \caption{Examples of rotation curves of other systems.}
    \label{fig:rotationcurves2}
\end{figure}

\noindent Some known exoplanetary systems \cite{ref11} or Earth’s system with its artificial satellites \cite{ref12} can also be explored using data available online. The number of available gravitational systems is thus more than sufficient for classroom implementation.

\medskip

\noindent The resulting plots exhibit the characteristic Keplerian behavior, as seen in Figure \ref{fig:rotationcurves2}, making it easy for teachers to verify whether calculations were performed correctly based on curve shape. If internet access is available, students may use online tools such as Google Sheets or Noragulfa's nPlot to compute and visualize their rotation curves and perform function fitting.

\medskip

\noindent Once the curve has been obtained, a function must be adjusted to the points. Some computational tools allow students to find that function. For the Solar System, the table shows the function found.
\bigskip

\begin{table}[H]
\centering
\begin{tabular}{|l|c|}
\hline
\textbf{Tool} & \textbf{Function} \\
\hline
Anthropic Claude & $v = 3.63 \times 10^7 \cdot R^{-0.49988}$ \\ \hline
OpenAI ChatGPT & $v = 1.03 \times 10^{10} \cdot R^{-0.50000}$ \\ \hline
Hangzhou Deepseek & $v = 9.75 \times 10^9 \cdot R^{-0.50000}$ \\ \hline
Desmos & $v = 1.03 \times 10^{10} \cdot R^{-0.49574}$ \\ \hline
GeoGebra & $v = 1.14 \times 10^{10} \cdot R^{-0.49980}$ \\ \hline
Google Spreadsheets & $v = 1.14 \times 10^{10} \cdot R^{-0.50000}$ \\ \hline
KDE LabPlot & $v = 1.01 \times 10^{10} \cdot R^{-0.49507}$ \\ \hline
Microsoft Excel & $v = 1.00 \times 10^{10} \cdot R^{-0.50000}$ \\ \hline
Noragulfa nPlot & $v = 1.03 \times 10^{10} \cdot R^{-0.49574}$ \\ 
\hline
\end{tabular}
\caption{Fitted power-law functions across various platforms (coefficients rounded to 2 decimals, exponents to 5)}
\end{table}

\noindent A pattern is exhibited here. Because the rotation curve is a function $v(R)$, the best function seems to be $v \propto R^{-1/2}$, which means

\begin{equation}
v^2 R = \text{constant} \label{prop}
\end{equation}

\medskip
\noindent It seems to be clear $v^2 R$ is a characteristic of every gravitational system. Doing this process for a particular system (like our Solar System) the constant is $1.2996 \times 10^{20} \ \text{m}^3 \text{s}^{-2}$. This process can be repeated for different gravitational systems like the outer planets in the Solar System.

\begin{table}[H]
\centering
\begin{tabular}{|c|c|c|}
\hline
\textbf{Central Body} & \textbf{Mass (kg)} & \boldmath{$v^2 R$} \textbf{($\text{m}^3 \text{s}^{-2}$)} \\
\hline
Sun     & $1.99 \times 10^{30}$ & $1.2996 \times 10^{20}$ \\ 
Jupiter & $1.90 \times 10^{27}$ & $1.25316 \times 10^{17}$ \\
Saturn  & $5.68 \times 10^{26}$ & $3.9204 \times 10^{16}$ \\
Uranus  & $8.68 \times 10^{25}$ & $5.86756 \times 10^{15}$ \\
Neptune & $1.02 \times 10^{26}$ & $7.25904 \times 10^{15}$ \\
\hline
\end{tabular}
\caption{Mass and $v^2 R$ values for selected central bodies in the Solar System.}
\label{tab:central_bodies}
\end{table}

\noindent In this table, five gravitational systems are listed with their characteristic $v^2 R$ values and the mass of the central body. If this relation is plotted (using a log-log scale due to the differences in data) a linear relationship appears and a linear regression can be applied.
\bigskip

\begin{figure}[H]
    \centering
    \includegraphics[width=0.43\textwidth]{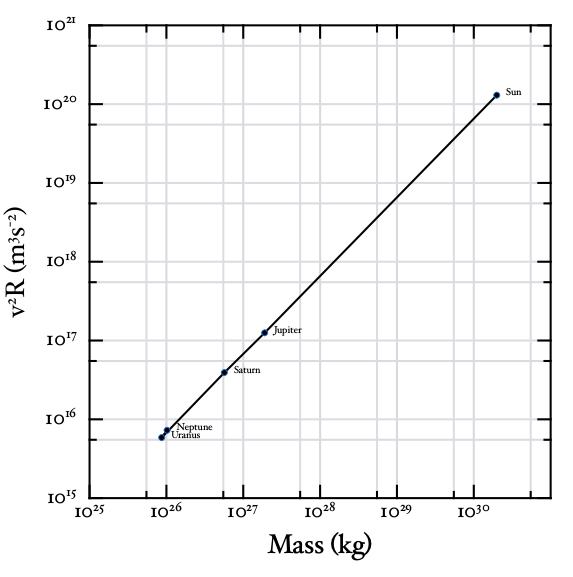}
    \caption{Linear adjustment of $v^2R$ versus mass of central body.}
    \label{fig:linearfit}
\end{figure}
\bigskip
\noindent  After that process, the obtained outcome for the slope is
 $ \Gamma = 6.54 \times 10^{-11} \ \text{m}^3\text{kg}^{-1}\text{s}^{-2}$. This suggests that the constant in \eqref{prop} is proportional to the mass of the central body. Consequently,

\begin{equation*}
v^2 = \frac{\Gamma \cdot M}{R}
\end{equation*}

\noindent and, due to the centripetal nature of gravitational attraction\footnote{In regular textbooks, the fact that the centripetal force is described by the equation $F = mv^2 / R$ is often demonstrated.}:

\begin{equation*}
F = \frac{\Gamma \cdot M \cdot m}{R^2}
\end{equation*}

\noindent This is the same expression found in \eqref{LUG}. The slope turns out to be the gravitational constant \( G \). The difference between this value and the accepted value \cite{ref15}  is by 2\%.


\bigskip

\section{Pedagogical Considerations}

\bigskip

\noindent Collaborative work is not only feasible in the classroom—it is a central component of this activity. While students may organize themselves freely, working in pairs tends to encourage more balanced engagement and shared responsibility. Once grouped, each team should be assigned a different gravitational system to analyze. Their investigation typically follows three main stages:

\begin{itemize}
    \item identifying rotation errors and adjusting the fitting function,
    \item calculating and reporting the $v^2 R$ constant along with the mass of the central body,
    \item and finally, determining the slope through linear regression.
\end{itemize}

\noindent These stages naturally support the development of key 21st-century skills, notably collaboration and teamwork, as emphasized by UNESCO \cite{ref16}. While independent work has its merits, working with others fosters discussion, constructive feedback, and richer interpretations. In particular, when students question results commonly accepted as axioms, they begin to appreciate the scientific process of building knowledge from real data. Within this environment, each team’s calculations evolve through conversation, analysis, and joint refinement—highlighting that deep understanding is cultivated through shared insight rather than solitary problem-solving.

\bigskip

\noindent Within this context, errors are not setbacks but learning opportunities. Some of the most common issues arise from inconsistencies in units, scaling inaccuracies in rotation curves, or incorrect data reporting. Students may also choose regression models that do not align with the expected theoretical behavior. For instance, including highly eccentric orbits—typical of comets or certain asteroids—can distort the curve and lead to misleading interpretations. Moreover, some software platforms may output exponents that deviate slightly from the expected $-0.5$, prompting students to question the model or propose alternative fits. These moments should not be met with rigid correction, but rather leveraged to promote critical discussion and deeper conceptual clarity.

\bigskip

\noindent Attention must also be given to the social dynamics of group work. Ideally, students with shared academic interests form teams that naturally collaborate well. However, inclusivity remains a priority. Students with disabilities or learning difficulties must receive the necessary support to engage meaningfully in the activity \cite{ref17}. While technology can be a valuable aid, allowing these students to work with paper-based calculations can also foster confidence and agency. Equally important is ensuring that each group contributes to the collective outcome, as the final regression relies on accurate and timely data from all teams. Although individual work is generally discouraged—due to the limited opportunities it offers for discussion—it should not be disallowed. Students who prefer to work alone may do so, provided their results are eventually compared with those of peers or the instructor, reinforcing the idea that collaboration is about dialogue, not forced social interaction.

\bigskip

\noindent Finally, when expanding the pool of gravitational systems, careful selection is necessary. As mentioned earlier, bodies with highly eccentric orbits—such as most comets and some asteroids—should be excluded because they are not in the line of the assumption of circular motion. Instead, instructors can look to satellite systems around Earth or exoplanetary systems with well-characterized orbits. Among the inner planets, only Earth provides sufficient satellite data for this activity. With intentional system selection and appropriate instructional support, this activity can remain inclusive and adaptable, offering students a rigorous yet accessible entry point into gravitational analysis.

\bigskip

\section{Discussion}

\noindent In any scientific activity, the way students present their results is a crucial part of the learning process. In this case, presentation choices depend significantly on available resources. Some groups may prefer to share their findings through simultaneous digital publication—especially when working with online tools—while others might work with offline software and opt for printed posters to display their outcomes. When internet access is limited or unavailable, scientific calculators or offline tools can still support function fitting and data visualization. Although this method has not yet been systematically tested for this activity, it offers a promising alternative for classrooms with restricted connectivity.

\bigskip

\noindent These presentations often initiate deeper discussions—especially around the topic of error. In astronomy and cosmology, measurement uncertainties are a natural part of the field. As students begin estimating the gravitational constant \( G \), they are prompted to reflect on how these values vary depending on data quality, system choice, and fitting techniques. This opens the door to a broader reflection on the nature of uncertainty in science: real-world measurements are rarely exact, and scientific progress depends on refining results through repetition, peer validation, and improved methods.

\bigskip

\noindent This rotation-curve investigation also adapts seamlessly to a Project-Based Learning (PBL) framework. Students might pose guiding questions such as, “How much does a planet weigh?” or “How can we measure the distance between Earth and the Moon?” These questions provide context and direction for mathematical exploration. Since the entire activity is algebra-based, it is accessible to students from eighth grade onward. Although this article does not explore conceptual adaptations in depth, the activity can certainly be reframed with a qualitative or discussion-based focus to meet the needs of more diverse learning groups.

\bigskip

\noindent One additional consideration concerns the use of theoretical tools like centripetal force. In many curricula, centripetal force is introduced as a pre-defined formula. However, within this activity, it can serve as a concept to be rediscovered rather than assumed. If the class has not yet studied this force in depth, instructors may benefit from a brief conceptual introduction or a quick experiment to ground its meaning \cite{ref18}. This context is particularly helpful when discussing planetary bodies with significant orbital eccentricity, such as Pluto, whose changing distance from the Sun complicates both the direction and magnitude of centripetal force. To address this issue, a useful strategy is to calculate the percentage difference between an orbit’s minimum and maximum radius. When the difference is small, it reinforces the assumption of near-circular motion—a key simplification that underpins this activity’s core calculations.

\addtolength{\textheight}{-12cm}   




\end{document}